# Aspects of SOA: An Entry Point for Starters

**Qusay F. Hassan**
**Faculty of Computers and Information Systems,
Mansoura University, Mansoura, Egypt**
qusayfadhel@yahoo.com

**ABSTRACT.** Because Service-Oriented Architecture (SOA) is one of the hottest topics that is currently gaining momentum, and the number of its adopters (both business and IT executives) is increasing in a tremendous manner, it is really a must to enlist important aspects related to it in order to allow these adopters to better understand the role that it can play in both software and business markets. These aspects varies from the definition of SOA and key components of it, different forms of support given by elite software vendors to it, its evolution history, the relationship between it and web services, the future expectations about its uses and benefits in different organizations, the relationship between SOA and Enterprise Application Integration (EAI), and various applications that can use it to overcome limitations related to other traditional methods. Moreover, challenges that face SOA in software market should be addressed and discussed in order to be able to see the big picture and to look for better solutions for them.
**KEYWORDS.** SOA, Service, Service Elements, Web Services, EAI, HAS, SOA Applications, SOA Predicts and Challenges, and SOSE.

## Introduction

Service-Oriented Architecture (SOA) represents a new paradigm that reflects a leap transition in both computing and software industries [T+06b]. It has emerged after decades of using distributed computing technologies to add a new element to software stack.

125

Abstractly, SOA is a computing paradigm that utilizes services as fundamental elements for developing systems. A service could be considered as a "set of software operations and components built in a way that allows them to be easily, flexibly, and dynamically integrated to cover both business and IT needs" [RH08]. Each service follows the client/server or request/response model in that it is expected to receive requests from clients to process them and finally forward back the returned result sets. The complexity of operations that might be performed by any service could vary from simple and short operations such as calculating simple formulas to very complex and long running business processes. Service-based applications are developed as independent sets of loosely-coupled services offering well defined interfaces to their users. These users can find and bind to services that they are interested in to start using them.

There is no question that SOA has gained a lot of traction last few years, and many of large computing vendors have moved to it including Microsoft, IBM [Chu05], HP, BEA, Sun Microsystems, SAP, Oracle, Cisco, and many others. Different kinds of support were given by these vendors in form of tools, languages, frameworks, and standards that emerged to sustain SOA systems. For example, one of main factors that led Microsoft to develop .NET Framework with great support to XML Web Services and Windows Communication Foundation (WCF) was to support SOA. Also, some new specifications have been provided by committees that are comprised of the aforementioned vendors to support XML Web Services yielding its next generation known as WS-*. These new specifications with no question have played a great role in maximizing the power of SOA and its inherent capabilities allowing better and higher adoption of it.

In addition to the adoption of SOA by large software vendors, it has been widely utilized by some of governmental agencies as a basic computing and modeling architectures including U.S Department of Defense (DoD) [Pau05].

Due to this great support to SOA, there is a big expectation that almost all computing units (both software and hardware) will be turned into services in the next few years to allow new applications and systems to be easily composed and recomposed to meet business and market ever-changing needs.

## 1. Evolution of SOA

In the past, many organizations were unsatisfied with the delivery of custom developed software applications because of their long development cycles, high costs, and inadequate output quality. Those reasons have led many of these organizations to buy large and packaged solutions such as Maintenance Resource Planning (MRP), Enterprise Resource planning (ERP), and Customer Relationship Management (CRM) systems. However, these solutions have not been good alternatives because they have come with new problems including very high costs, difficult implementation, integration complexities, long customization cycles, and limited flexibility.

Certainly, two key pressures usually affect software market listed as follows [E+04]:

1. **Heterogeneity:** Most of enterprises today have a giant mix of systems that were built over time using a verity of technologies coming from different sources. Integrating such products was always a nightmare for IT executives.
2. **Change:** The rate of change is increasing in an aggressive manner especially after the emergence of globalization and e-business. This high rate of change represents one of the hardest problems that face organizations and enterprises in market. It also leads them to face fierce competition to stay in market. To solve this challenge, software cycles should be shortened and information should be offered in a quick manner in order to be able to compete and gain high revenues.

According to the aforementioned pressures, software applications have turned over years from large and complex packaged (siloed) systems such as inventory management, finance, and personnel systems to the famous client/server model that was designed to support cross-functional business processes. Although the client/server systems were very beneficial to organizations, they were still monolithic.

IT executives think that SOA may enable them to alleviate many of problems related to heterogeneity, interoperability and market ever-changing needs by allowing them to leverage existing IT investments in a more efficient form to be able fulfill organizations' goals effectively.

It is worth mentioning that the history of SOA goes back to a concept known as software-as-a-service (SaaS) which has first appeared with Application Service Provider (ASP) software model [GW02].

Simply, ASP is a "third party entity that deploys, hosts, and manages access to a packaged application and delivers software-based services and solutions to customers across WAN from a central data center" [GW02], [KBS04]. So, this ASP is responsible for managing, updating, maintaining, and supporting hosted applications as well as underlying infrastructures. These hosted applications are delivered over network on subscription or rental basis.

Unfortunately, ASP model has suffered from several inherent limitations such as inability to provide complete customizable applications that resulted in a new generation of monolithic and tight-coupled architectures. These limitations have allowed the SOA paradigm to emerge to offer delivery of complex business processes in form of network addressable components known as services that could be accessed and reused everywhere by everyone on condition that access permissions are granted to requestors. This model enabled end-to-end integration between different systems or even organizations with the ability to construct new applications and business processes on the fly to meet new and unexpected business needs.

## 2. What is Service?

Basically, service is a self-contained (offers different functionalities related to one business or technical area/sub-area), cohesive (all related functionalities are placed together), black box (consumers know nothing about its internals, and underlying technologies) software component that encapsulates a high level business/technical concept that can cover specific area of whole system.

A service could be designed to be *fine-grained* or *coarse-grained*. A fine-grained service can handle small and specific functionality such as log service, database access service, fees calculation service, interest rate calculation service, etc. A coarse-grained service is more powerful service that is usually composed of a number of fine-grained services to handle more complex and wide range of related functionalities such as loan management service, purchase order service, insurance claim service, etc.

Moreover, a service might be *stateful* or *stateless*. A stateful service retains and manages service states during its execution and between different requests, for example, a service that is responsible for handling loans must be stateful in order to keep state information about the loan being processed for each bank client. Reversibly, a stateless service does not retain its states between different invocations. The stateless service is the most used type of services in SOA because it allows loose-coupling between requestors and offered services, enabling wider range of clients to use these services.

Also, a service might be *short-lived* or *long-lived*. The execution of transactions offered by a short-lived service can take sub-seconds or seconds to finish working and return back generated results, for example, getting or updating a database record such as customer information. On other hand, the execution of long-lived service may take minutes, hours, days, or even months to reach its final state, for example, a loan management service may take months to grant a new loan to one of bank customers.

## 3. Service Constituent Elements

Technically, any service consists of three main parts listed as follows [KBS04]:
- **Contract:** It provides both *formal* and *informal* specifications of service. Formal specifications use one or more of available description languages such as IDL and WSDL to describe information related to technical areas of service such as underlying programming language(s), middleware(s), network protocol(s), and other runtime aspects. On other hand, informal specifications are textually presented to provide general information such as the purpose, functionality, constraints, usage of exposed service, and expected response time.
- **Interface:** It provides technical representation of service operations that are available to be invoked by clients. Any interface may contain information about public operations, parameters, and return types. These interfaces are also known as *stubs* or *proxy classes*.
- **Implementation:** It contains actual logic of service that might be related to accessing data, business logic, etc. This implementation logic could be encapsulated internally within service itself or it may be provided by other external artifacts such as other programs, code libraries, components, legacy systems, etc.

## 4. SOA in Context of Object-Oriented Programming and Component-based Development

Service is different than other software modules in that it tends to provide and execute technical/business functionalities and processes. Thus, service is an abstract, sophisticated, and coarse-grained processing unit that wraps sets of software components and objects working together in a harmonic environment to provide functions that the service represents.

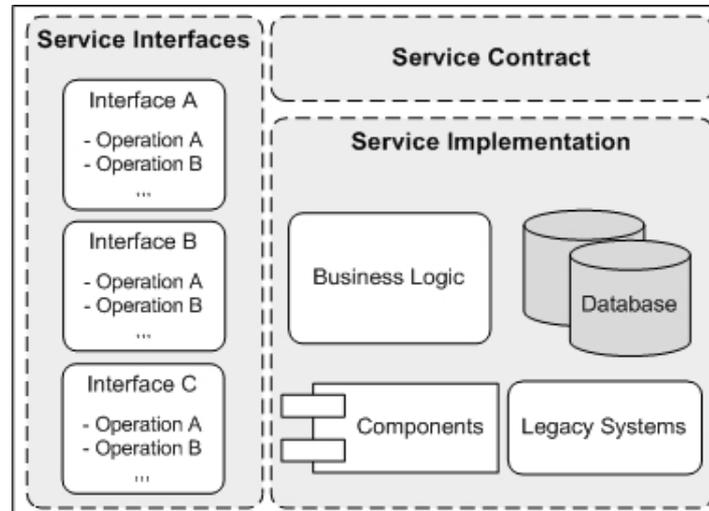

**Figure 1:** Essential Service Elements

There is no question that one of the most important advancements in modern software development methodologies is the evolution of Object-Oriented Programming (OOP). This methodology depends on objects to allow programmers to model software problems in a real-world manner by mapping software terms to entities from our lives. Certainly, any object can have only states (properties) that describe its characteristics and distinguish it from other objects, and behaviors (operations) that are responsible for changing state values [Boo90]. Although OOP is a great way to model software systems, objects suffer from a number of limitations including being very fine-grained elements, as each object only represents one small entity or even a part of a more complex entity. Moreover, objects are tightly-coupled due to high dependency on inheritance that allows programmers to extend one or more objects with new features. These limitations made the degree of dynamicity and reusability very small, as it only allows programmers to instantiate defined classes and define relationships between them into code at design-time (before compilation-time).

Limited reusability offered by the OOP model has forced software execs to look for a more reusable form. This was achieved by the emergence of the Component-based Development (CBD) model that utilizes components as main elements for building software systems. Component is a more complex and abstract term than object, as it represents a unit of composition of groups of objects that work together to provide needed functionalities, allowing programmers to access them through a set of contractual interfaces [Szy03].

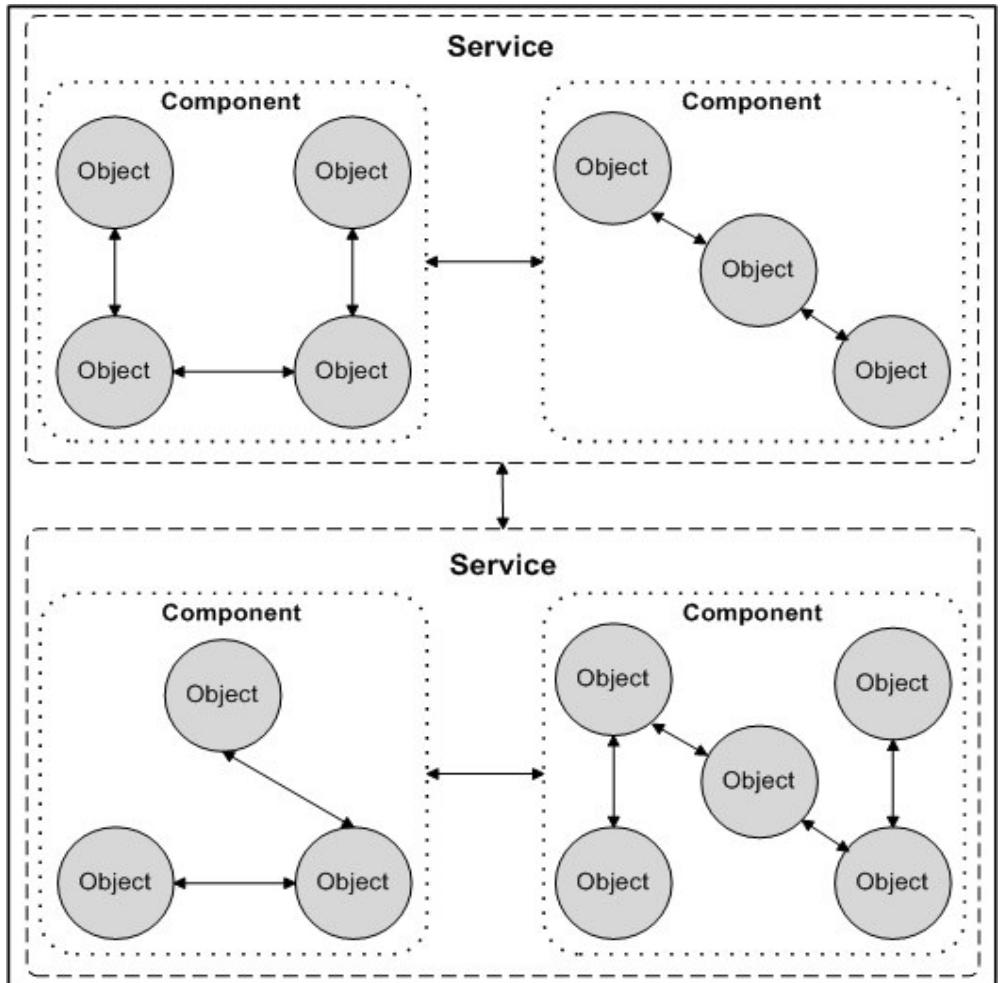

**Figure 2:** Service in the context of Components and Objects

Technically, SOA adds a layer of abstraction on top of CBD in a way that allows programmers to use and combine components built with different programming languages, frameworks, and technologies. Also,

131

service-orientation supports features such as dynamic discovery of services, loose-coupling between different services, and improved level of reusability. And that is why, services are closer to the concept of business transactions allowing service consumer to be totally unaware of technical issues related to the underlying components as long as defined services offer needed functionalities at expected service-level agreement (SLA).

## 5. SOA and XML Web Services

Although many technologies could be utilized to implement SOA including CORBA; message queuing technologies (such as JMS and MSMQ); RMI; RPC; and COM/DCOM/COM+, XML Web Services technology is the most favorite choice for its adopters due to a number of factors including:

- **Standards-based:** XML Web Services technology has emerged as a result of cooperation between different large software vendors including Microsoft, IBM, Sun Microsystems, Oracle, SAP, BEA, and many others. This has given XML Web Services a great support in almost all modern programming languages, frameworks, and tools.
- **Dynamic:** Because different tools and programming languages could be used to construct the logic of XML Web Services, they enable faster and more efficient implementation for applications that require high dynamicity in response to market circumstances and business requirements such as dynamic e-business.
- **Modular:** It is modular in nature, so, it allows its adopters to encapsulate business logic in terms of modules that could be deployed and used separately by available requestors.
- **Composable:** Simple XML Web Services might be aggregated to yield more complex ones. This allows complex business processes to be easily designed, implemented and modified.
- **Simple:** Implementing SOA using XML Web Services is very easy if compared with other distributed technologies such as CORBA and COM family.
- **Cheap:** It is a much cheaper than other technologies especially the proprietary ones.

    Additionally, the basic models of SOA and XML Web Services are almost identical, as they both are composed of similar parts listed as follows [Pap03]:

- **Service Provider:** It is responsible for creating, managing, and maintaining needed services. It is also responsible for providing all information needed to describe offered services. In XML Web Services, parameters passed to any web service and results returned from it are represented as XML (extensible markup language) documents.
- **Service Consumer:** It is an entity that is interested in a specific service. Service consumer is responsible for finding needed services and binding to them in order to use them. In XML Web Services, communication between service consumer and service provider is mainly performed using SOAP (Simple Object Access Protocol) messages.
- **Service Registry:** It is a special type of databases that allows service provider to publish available services with all related contracts and metadata information. UDDI (Universal Discovery, Description, and Integration) is the main type of registries used in the world of XML Web Services allowing service consumers to query published services either programmatically or through some GUI screens to find those that best fit their needs.
- **Service Contract:** It holds information that describes available services. When a service provider wants to expose one new service, he must provide its contract to service registry in order to allow service consumer to find it and to know all information and guidelines that allow him to use it. WSDL (Web Service Description Language) is the description language that is used by XML Web Services to describe different information (both functional and non-functional) about available services including their URLs, ports, protocols, operations, parameters, contact information, etc.

## 6. SOA Benefits

In fact, SOA is a fruitful approach that offers lots of benefits and advantages to its adopters and practitioners including:
- **Loose-Coupling:** Because consumers can access services in runtime through set of contractual interfaces rather than static dependencies that are defined in code, then SOA allows consumers to dynamically compose and recompose different services to build new (composite) applications. In addition, SOA allows business execs to orchestrate available services in ways that allow them to build new business processes or even modify existing ones in order to meet business requirements.

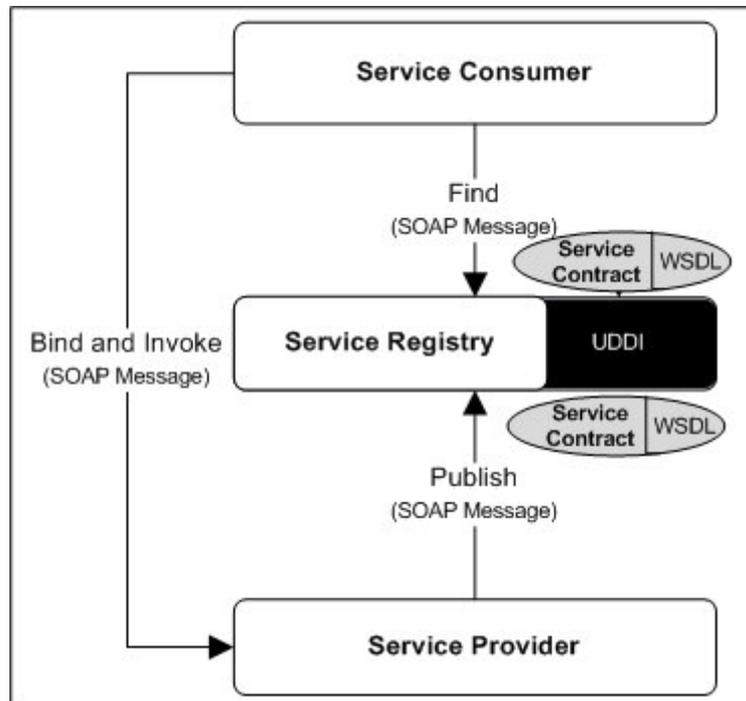

**Figure 3:** SOA in Context of XML Web Services

- **Location Transparency:** With SOA, there is no need for consumers to know physical paths of used services as they can only access them through set of public URLs/IPs defined in service registries just like online websites. This feature totally exempts consumers from awareness of technical complexities such as hardware, operating systems, frameworks, and communication infrastructures used to host services. Moreover, it enables service providers to change physical paths or offered services whenever needed without any impacts on service consumers.
- **More Reusability:** Once a service created, it can be reused more and more by different consumers and in various applications. Moreover, services might be composed into other services for building more complex ones that are responsible for executing sophisticated business processes. To enable reusability of a service, it should be created at the optimum *granularity* and *coherency* levels. Granularity refers to size and scope of the service, whereas, coherency refers to putting related (technical and/or business) functionalities together in one service.
- **Higher Productivity:** If developers reused available services, then projects can go faster, and same development team can be assigned to work on more projects, resulting in great save in total development costs.
- **Leveraging Legacy Components:** Enterprises usually have lots of legacy components and systems that were built overtime to cover different needs. Utilizing these components in modern solution may not be easy or even possible due to technical and cost constraints. However, SOA could allow enterprises to make use of these assets by wrapping them into a form that might be used by modern solutions.
- **Greater Interoperability:** SOA came with set of guidelines and standards that enable different enterprises/organizations to be integrated together at lower costs and less efforts, and with higher effectiveness.
- **Higher Agility:** SOA promotes high levels of agility by making systems easier to be built and modified. This agility could be gained from avoiding redundant and isolated systems, and instead, creating well integrated systems that fulfill shared needs between different departments in enterprise/organization or even between different enterprises/organizations and joint-venture (JV) companies that may work together.
- **Better Alignment between IT and Business Execs:** One of common drawbacks of traditional software methodologies is inefficient communication dialogues between business execs and IT execs. This drawback makes the creation of needed software components a hard mission to accomplish. On other hand, because SOA defines both business requirements and software functions as services, then it offers a better communication dialogues between IT and business representatives, allowing each of them to give inputs that are mostly understood by each other.

## 7. SOA Predicts

According to the importance of SOA, many of market and research organizations have predicted that SOA will change both business and software market in an explosive manner. For example, Gartner Group has announced that SOA will be used in part in more than 50% of business applications in 2007, and it has expected that it will be used in 80% software systems by 2010. The IDC published a report numbered #27093 in which it predicted that projects that use web services would have $7.1 billion by 2006 with an annual compound rate at 116% per year in the next few years. The Radicati Group has an expectation that the SOA market would reach $6.2 billion by 2008 with an annual compound rate of 50%.

## 8. SOA and Enterprise Application Integration

Typically, integrating different systems together is one of the well known uses of SOA [Lin03]. This can be accomplished by different methods including:

- **Service-Oriented Integration (SOI):** It allows integration of different systems by means of services. SOI is different from traditional integration methods in that it does not depend on information sharing that suffers from Extract/Transform/Load (ETL) method which enforces each requestor to build his own logic in form of proprietary interfaces to extract and transform accessed data from available data sources, but rather, it is based on exposing these information in form of readymade services that could be reused over and over by different requestors in various applications. In fact, SOI acts as a second generation of the Enterprise Application Integration (EAI), as it allows services to be shared and reused in different systems. This software phenomenon is also known as composite applications. Simply, composite applications term refers to ability to quickly construct new systems by assembling/reassembling available services.
- **Business Process Modeling (BPM):** Briefly, it tends to integrate processes provided by different enterprises to allow end-to-end integration. This could be easily done if these processes are built in form of coherent services. Applying SOA concepts to build business processes in form of services allows business execs to easily orchestrate these services using BPM and workflow tools without extensive technical knowledge.
- **Portal-Oriented Integration (POI):** It tends to offer services that comprise logic needed by various portals, for example, Yahoo and MSN provide end-users with various types of information including information about weather, news, gossips, sports, etc. These varieties of information are usually gathered from other organizations that act as service providers for wide range of consumers. Also, Web Service for Remote Portlets (WSRP) which is imposed by SOA and web services technology to allow integration from presentation-based perspectives could be utilized to display section(s) of remote portals into other portals in a very simple way and with almost no efforts [Sch02]. Certainly, these approaches are really good for most of e-business forms including both Business-to-Business (B2B) and Business-to-Consumer (B2C) methods as it allows meeting for business needs in easy and fast manner.

## 9. SOA and Hardware Industries

It is good mentioning that many hardware industries have supported SOA in the designs of their products. For example, Intel is applying service-oriented approach to system design where they treat Hardware as Services (HAS) [CHC06]. This concept aims to design hardware units as services in order to be able to interoperate them with any other resources. This feature could encourage other important terms such as virtualization which is widely used in grid computing to enable sharing of resources such as files, computers, servers, mainframes, supercomputers, software components, and underlying data in a very large scale and in heterogeneous environments to allow these resources to collaborate to solve complex computation problems [JF04].

## 10. SOA Applications

Nowadays, SOA is being widely adopted in many various applications inside enterprises that serve different purposes including:
- **Finance and Trading:** Financial and trading section represents one of the most important fields that could utilize SOA to gain its inherent advantages. Typically, SOA could be leveraged in different forms in financial applications such as building, enhancing, or even integrating them together through a standardized set of interfaces [D+05]. Different financial systems could benefit from SOA capabilities such as banking systems [RHQ08], Supply-Chain Management (SCM) systems [Z+06], etc.
- **E-Business:** Applying SOA in e-business aims to add more dynamicity to participating systems to be able to add, remove, modify, or integrate them in a more flexible and efficient manner [Bih06]. Clearly, SOA could be considered as the next generation of integration methodologies that could be used in e-business field, as it could enable different organizations to share information to allow merger and acquisition [AKS06]. SOA has overcome problems related to e-business field that have emerged as a result of using expensive and less effective technologies such as proprietary interfaces, ETL, and Electronic Data Interchange (EDI). Also, new standards and technologies that have recently emerged to sustain SOA in e-business such as BPEL4WS have allowed different organizations and enterprises to share their business processes with others, enabling them to have end-to-end integrations [Pas05].

- **E-Health:** For the last two decades, the IT executives specialized in providing healthcare systems were trying to manage the costs and quality of provided systems. However, these systems have suffered from being extremely decentralized and scattered. This problem has prevented them from developing a single system that meets all their needs. Regarding to this problem, many research projects have been made to utilize SOA in e-health field, and as a result, many experiments have been conducted and many research papers have been published to discuss and solve these issues [OB06].

- **E-Government:** One of the most important demands that should be fulfilled in e-government applications is that they should be based on modern and widely accepted standards [O+03]. SOA and web services could be combined together to design and build interoperable, reliable, dependable, and secured e-government applications. This could be accomplished by dividing functionalities provided by e-government applications to a set of services including administration services, infrastructure services (such as security, printing, notification, message transformation and mediation), processes and sub-processes services, and user interface services. These different services could be composed and recomposed to meet different requirements whenever needed [M+05].

## 11. SOA Challenges

It is worth mentioning that one of the most important challenges that face SOA is the lack of knowledge about terms and aspects related to it [Lap05]. Most of software courses offered by software faculties are only based on traditional methodologies such as OOP with some related topics such as Unified Modeling Language (UML), Object-Oriented Analysis and Design (OOAD), and Object-Oriented Languages. To solve this problem, curriculums offered by software faculties should be extended to contain different aspects of SOA including definition and types of services, approaches and strategies, enabler tools and technologies, the role that SOA can play in integrating different systems, and missing points in current SOA models. This with no doubt can play a great role in yielding a new generation of software specialists that can meet new market needs.

## 12. Service-Oriented System Engineering

As mentioned, SOA is a new paradigm in IT market, so, applying it in an effective and efficient manner really needs some different software engineering techniques. Service-Oriented System Engineering (SOSE) is an emerging topic that comprises set of software engineering techniques needed for SOA analysis, modeling, specifications, creation, testing, debugging, monitoring, and governance [T+06a]. The main focus of SOSE is to create reliable, secure, dependable, and trustworthy service-oriented systems, and that is why many conferences and workshops have been held to cover these needs including IEEE International Symposium on Service-Oriented System Engineering (SOSE), International Service Availability Symposium (ISAS), International Workshop on Collaborative Computing, Integration, and Assurance (WCCIA), the International Symposium on Service-Oriented Applications, Integration and Collaboration (SOAIC), and the International Workshop on Service-Oriented Software Engineering (IW-SOSE). All these conferences and workshops have extensively discussed main concerns related to SOA, and many papers related to this topic have been published.

## Conclusion

In this paper we have discussed some of important aspects related to SOA varying from an introduction to its terms, benefits, and challenges to different applications that could utilize it. Certainly, these points aim to clarify the role that SOA can play in both software and business markets in the coming few years. Additionally, an introduction to Service-Oriented System Engineering has been given to point out its relationship with SOA in order to put its adopters on the right track.